\documentclass[twoside]{article}
\usepackage{fleqn,espcrc2}
\usepackage{epsfig}
\usepackage[figuresright]{rotating}

\usepackage{graphicx}
\usepackage[figuresright]{rotating}

\newcommand{\AmS}{{\protect\the\textfont2
  A\kern-.1667em\lower.5ex\hbox{M}\kern-.125emS}}

\title{Development of a radiation hard version of the Analog Pipeline Chip APC128}
     
\author{
        Michael Hilgers
         \address[IPP]{Institute for Particle Physics, ETH Z\"urich, CH 8093 Z\"urich, Switzerland}
 \thanks{Corresponding author.  Fax: +41 1 6331233, Tel.: +41 1 6334009, e-mail: hilgers@phys.ethz.ch },     
        Roland Horisberger
         \address{Paul Scherrer Institut (PSI), CH 5232 Villigen-PSI, Switzerland}}         
\begin{document}

\begin{abstract}
The  Analog Pipeline Chip (APC) is a low noise,
low power readout chip for silicon micro strip detectors
with 128 channels containing an analog pipeline of 32 buffers depth.
The chip has been designed for operation at HERA with a power 
dissipation of $300-400\, \mu \mbox{W}$ per channel and has been used also in 
several other particle physics experiments. 
In this paper we describe the development of a radiation hard version 
of this chip that will be used in the H1 vertex detector 
for operation at the luminosity upgraded HERA machine.
A 128 channel prototyping chip with several amplifier variations  
has been designed in the radiation hard  DMILL technology and measured. 
The results of various parameter variations are presented in this paper.
Based on this, the design choice for the final production version of 
the APC128-DMILL has been made. 
\vspace{1pc}
\vspace{1pc}

\vspace{1pc}
\end{abstract}

\maketitle

\section{Introduction}
Since approximately 15 years silicon vertex detectors 
are operated successfully in high energy particle physics experiments.
The small pitch of  silicon micro strip detectors implies 
a very high channel density for the readout electronics
which is normally achieved with a dedicated CMOS readout chip.
The Analog Pipeline Chip (APC) is a  readout chip 
primarily designed for the Central Silicon Detector (CST)
\cite{cstpaper}
of the H1 experiment 
\cite{h1detector1,h1detector2}
but furthermore successfully used by other experiments
\cite{bstdevelop,hermes,bruessel,na47}.

The basic functionality of the existing chip is described 
in section \ref{apcgeneral}.
As explained in section \ref{apcfailure}
radiation damage of the chip was observed in the H1 experiment
motivating the design of a new radiation hard version of the APC.
We report  in section \ref{apcprototype} on an 128 channel APC prototype 
designed and produced in the radiation hard DMILL technology.
Five different amplifier configurations implemented on this prototype 
are explained.
The measurements made with the different amplifiers are presented in
section  \ref{apcmeasure}.
The conclusions we drew led to the final design that is currently 
in production (section \ref{dmilldesign}).

\section{\label{apcgeneral} General description of the APC-chip}
\begin{figure*}[t]
\begin{center}
\leavevmode
\epsfig{file=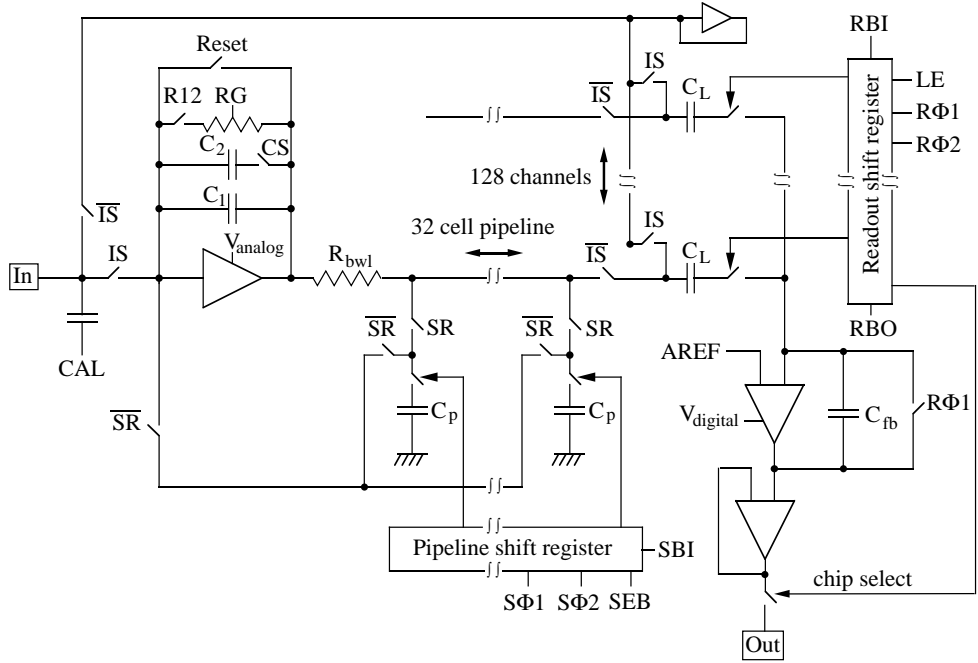,width=13cm}
\end{center}
\vspace{-1cm}
\caption[]{\label{apcschematic}Electrical diagram of the Analog Pipeline Chip }
\end{figure*}

Figure \ref{apcschematic} shows schematically 
one of 128 channels of the Analog Pipeline Chip . 

Each channel has a charge sensitive, low noise, low power preamplifier 
followed by a 32-cell storage pipeline.
The storage pipeline consists of switched capacitors.
When disconnecting the capacitors they store a charge proportional to the 
output voltage of the preamplifier. 
The storage capacitors of the pipeline are controlled by the pipeline shift register.
This pipeline shift register 
can be operated with a frequency of 10.4 MHz
corresponding to the HERA bunch crossing rate. 
An analog voltage of $\approx 2\, \mbox{V}$ is exclusively supplied 
to the preamplifier 
whereas all other building blocks are 
operated with $5\,\mbox{V}$ supply voltage.    

To read out the pipeline (e.g. in case a trigger decision is made
by another detector component)
the sampling is stopped and the input of the 
charge sensitive preamplifier is disconnected from the sensor by the $IS$ signal 
(see Fig. \ref{apcschematic} ).
By means of the $\overline{SR}$ signal the pipeline 
is connected to the input of the preamplifier.
The bit in the pipeline shift register determines now, 
which storage cell is re-read by the amplifier.
This  so called re-read architecture, 
a peculiarity of the APC chip, 
offers various advantages:
As the preamplifier re-reads its own signal a perfect matching of 
the operating point is automatically given. 
The ratio between the pipeline capacitor $C_p$ and 
the feedback capacitor $C_1$ or $C_{1}+C_{2}$ allows to define a signal gain while 
re-reading.
Reading the charge from two (or more) pipeline buffers in parallel
or consecutively allows to sum up these pipeline buffers.

The re-reading preamplifier writes the signal to the latch-capacitor $C_{L}$
where it is stored until readout.
This latch-capacitor provides the possibility to subtract different buffer 
of the same pipeline
The possibilities to do on chip analog signal processing
have been studied in an early 12 channel prototype version \cite{apcsac2}.  
Moreover the ratio between $C_{L}$ 
and the feedback capacitance of the charge sensitive readout amplifier   
($C_{fb}$ in Fig. \ref{apcschematic})
form another gain stage before the signal leaves the chip.
Finally the latch-capacitor $C_{L}$ decouples the (very different)
 working point of the readout amplifier and the preamplifier.

The serial readout of the 128 latch-capacitors of the chip 
is controlled via the 
readout shift register that connects one channel at a time to a 
common two stage readout amplifier. 
Both the re-reading architecture and the serial readout are area-efficient 
concepts implicating that most of the area on the APC chip is occupied
by the bond pads and the pipeline capacitors.
This leads to a total chip size of $6.3\, \mbox{mm}$ by $3.5 \mbox{mm}$. 

\subsection{The preamplifier}
The preamplifier consists of a n-MOS transistor and a  
p-MOS transistor with a large W/L ratio forming 
a push-pull stage.
In this inverter configuration the devices act as mutual loads to each other.  
This circuit offers maximum transconductance and therefore 
minimal white channel noise contribution  at minimal power dissipation.

Figure \ref{apcschematic} shows the use of the inverting amplifier 
in a charge sensitive configuration.
The feedback capacitance $C_1$ is given by the parasitic gate-drain 
capacitances of the large input transistors.
An additional feedback capacitor   $C_2$
can be connected to enlarge the  feedback capacitance.  

A reset switch allows a fast discharge of the amplifier. 

A constant and slow discharge of the charge sensitive amplifier is done 
by means of the feedback resistor $RG$.
This resistor is realized as a n-MOS transistor with a small W/L ratio. 
The value of the resistor and therewith the discharging time constant 
 can be adjusted by the gate voltage of this n-MOS transistor ($RG$).
The discharge resistor is disconnected during the re-read procedure 
by the $R12$ switch. 
In this mode the preamplifier is fully integrating.

Moreover this resistor performs a second vital task. 
Since in the CST the sensors are DC-coupled to the APC-chip
the resistor must conduct the leakage currents of the silicon strips 
to the output of the preamplifier where the current is absorbed.
Due to the voltage drop across the feedback transistor 
this mechanism limits the maximum value of the resistance.
The preamplifier can tolerate a voltage drop of several hundred  mV between 
input and output 
\footnote{The preamplifier individually can in fact tolerate voltage drops up to 
roughly 300 mV. But summing up three pipeline buffers as done in H1 effectively reduces this value 
to $100\ \mbox{mV}$.}.
Expecting maximal leakage currents of $100\, \mbox{nA}$ per strip 
implies a feedback resistance of less than $1\, \mbox{M} \Omega$.

\section{\label{apcfailure} Observations of radiation damage to the SACMOS-APC}
For the readout of the CST vertex detector of the H1 experiment 
640 chips, fabricated in the SACMOS-1$\mu$ technology, have been used.
After three years of successful operation of the CST detector 
observations of radiation damage to the APCs were made:
\psfull
\begin{figure}
\begin{center}
\leavevmode
\epsfig{file=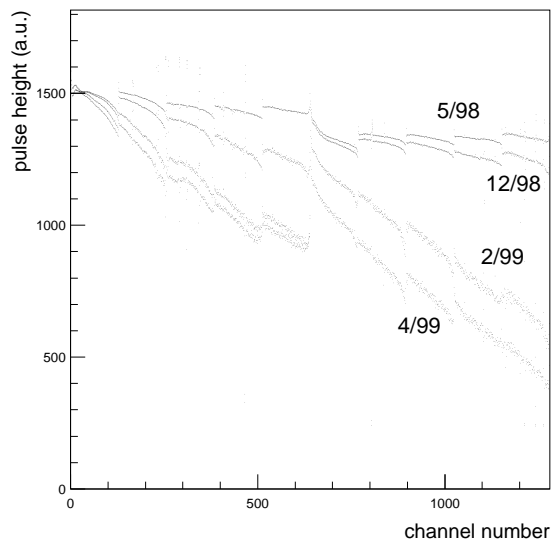,width=7.8cm}
\end{center}
\vspace{-1.cm}
\caption[]{\label{apcschaden}Four sets of pedestals of the SACMOS Analog Pipeline Chip in the CST-Detector at H1 indicating radiation damage of the chips
as a function of date from May '98 to April '99. 
The channel number is proportional to the time the channel waits until 
it is read out (see text for details). }
\end{figure} 

Figure \ref{apcschaden} shows the pedestals of a serial readout 
of 1280 channels (10 chips of 128 channels) of the inner layer of the CST. 
The pedestals are the average channel voltage  (in arbitrary units) 
with no signal present.
As the sequential readout is done with a clocking frequency of 1.5 MHz 
the last channel 
(rightmost in Fig. \ref{apcschaden})
waits $1280 \times 667 \, \mbox{ns} \approx  1\, \mbox{ms}$ for its readout.
For undamaged chips the pedestals have a level 
that is almost independent of the channel number 
respectively the waiting time.

As radiation damage proceeds 
it can be seen in figure \ref{apcschaden} 
that high channel numbers show systematically lowered positions.  
This effect gets more and more pronounced as longer the chips 
are exposed to the radiation environment of the experiment.
In the  outer layer of the CST, 
where the radiation dose is considerably smaller, 
the change of the pedestals is therefore consistently smaller
(by a factor of $\approx 1/3$).

The change of the pedestals with ongoing time 
is explained by increased radiation damage 
( until 5/1998 $\approx 100 \, \mbox{Gy}$ total dose, 
  until 4/1999 $\approx 300 \, \mbox{Gy}$ total dose)
that accelerates the discharge of the signal storing capacitors 
$C_L$ and $C_1$.
The discharging currents on the chip are either 
a result of the  sub threshold leakage currents of the n-MOS transistors 
that are used as switches (see figure \ref{apcschematic})
or due to a reduced device isolation.  
         
In principle pedestal shifts do not disturb the functionality of the system 
as long as they stay within the dynamic range of the readout chain.
However in future HERA will run with higher beam currents and 
stronger focusing at the interaction point \cite{H1upgrade} 
and a higher radiation exposure of the CST is anticipated and therefore 
it is not guaranteed that the present front end chips will stay functional.
This reasoning led to the conclusion that radiation hard analog pipeline chips 
are vital for the future data taking of the H1 Central Silicon Detector.

\section{\label{apcprototype} Prototyping of the DMILL-APC}
The basic requirement of the new radiation hard APC is a strict compatibility 
with the existing readout and power supply system of the CST 
\cite{cstpaper,optical}.
Furthermore this allows also other projects using the
SACMOS-APC a smooth change-over to the usage of the APC128-DMILL. 

As a prototype chip a complete 128 channel version 
with five different amplifier configurations
has been designed
in the $0.8\, \mu$ DMILL-BiCMOS-technology 
\cite{dmill,rd29} offered by TEMIC/MHS \cite{temic,temicaddress}. 
The various amplifiers were also realized as test structures 
to allow direct performance measurements without using the complete
pipeline mechanism. 

Since it is not self-evident that an analog circuit working properly in one technology can be transferred directly to other technologies,
we realized different layouts for the preamplifier, 
in order to identify a design with optimized
low noise performance at low power.

This procedure proved to be very valuable as it turned out 
that a direct one-to-one translation of the SACMOS APC preamplifier 
did not work  satisfactory in the DMILL version.  

\subsection{Digital part}
The digital logic of the new DMILL-chip was realized 
as similar to the SACMOS-chip as possible. 
The readout shift register (see section \ref{apcgeneral})
is a dynamic shift register
whereas the pipeline shift register is a static cell based 
on cross-coupled inverters.
The area requirements for the different functional blocks 
was only marginally larger in the DMILL technology.

\subsection{Preamplifiers}
The five different amplifier configurations contain three variations of the 
single stage amplifier as used in the SACMOS version 
and a pair of two-stage amplifiers.

\subsubsection{Single stage amplifier \label{singlestage}}
The three versions of the single stage amplifier differ only by 
the channel length L. 
We chose L= $1\, \mu \mbox{m}$, L= $2\,  \mu \mbox{m}$, 
L= $3\, \mu \mbox{m}$. 
Common to all designs 
was a channel width of $W=620\, \mu \mbox{m}$ for the p-FET 
and $W=290\, \mu \mbox{m}$ for the n-FET.
These width parameters are roughly $15\%$ smaller  than the original 
SACMOS parameters.

The translation of the  long feedback transistor of the SACMOS APC 
( $ \mbox{W/L} = 1.4\, \mu \mbox{m} / 368\, \mu \mbox{m}$)
required a splitting into five n-MOS transistor in series 
with a W/L ratio of $2.2\, \mu \mbox{m} /90\, \mu \mbox{m} $ each.

The additional feedback capacitance ( $C_2$ in figure \ref{apcschematic})
was designed to be  $540\, \mbox{fF}$. 

\subsubsection{\label{twostagedesign} Two stage amplifier}
Channels with a two-stage amplifier scheme were designed in order to  
drive the capacitive load of the analog pipeline.
Moreover this pipeline buffer stage was designed to give a further 
amplification of a factor of two 
with the aim of reducing the noise contribution 
coming from the pipeline switching.

For the channels with two stage configuration 
the first stage was identical to the single stage 
preamplifier with a channel length of $ L=3\, \mu \mbox{m}$.

The pipeline buffer stage consists of a 
simple differential transconductance amplifier.
The differential pair is made of two p-MOS transistors. 
A n-MOS current mirror act as a load.

With a p-MOS current source a typical operating current of $20\, \mu \mbox{A}$ 
per channel is defined.

\psfull
\begin{figure}[thb]
\begin{center}
\leavevmode
\epsfig{file=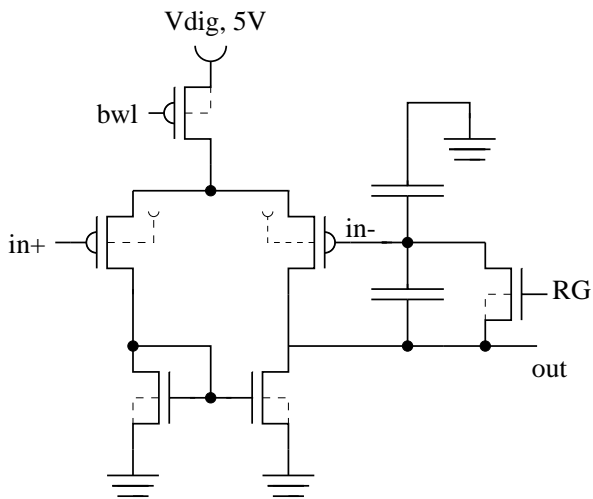,width=7.8cm}
\end{center}
\vspace{-1.cm}
\caption[]{\label{wilson1}Simple transconductance amplifier with feedback loop as described in the text.}
\end{figure} 
The feedback loop connects the output of the second stage 
with the negative input of the differential pair 
forming a mere follower for DC signals. 
AC-signals undergo an amplification defined by the capacitive divider in the feedback loop. 
Both capacitors are chosen to be 1pF resulting in gain 2.     
Figure \ref{wilson1} shows the circuit.
The positive input of the transconductance amplifier 
is directly connected to the output of the preamplifier.
 
In one variation the feedback resistor is a simple n-FET 
with small W/L-ratio.

\begin{figure}[thb]
\begin{center}
\leavevmode
\epsfig{file=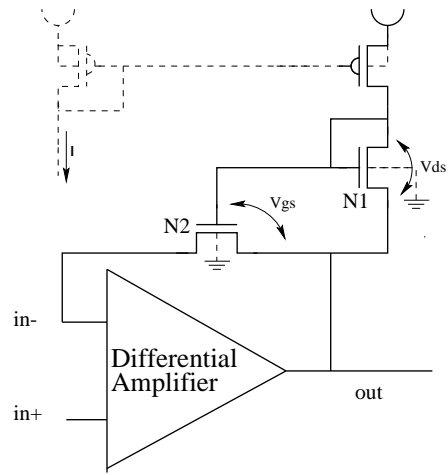,width=5.8cm}
\end{center}
\vspace{-1.cm}
\caption[]{\label{wilson2}Feedback configuration of second prototype variation
for the pipeline buffer stage,  
see text for details. 
The capacitive divider is omitted for clarity. 
}
\end{figure}

A second variation of this two-stage amplifier uses
a  somewhat more complicated version of the feedback transistor:
Instead of having a n-MOS transistor with a fixed gate voltage as 
resistive element (see figure \ref{wilson1}) 
this n-MOS transistor was implemented in such a way that
the gate voltage follows the amplifier output swing, 
thus keeping the gate-source voltage fixed  \cite{blanquart}. 
As the gate voltage follows the amplifier output swing we refer 
to this circuit  as a riding feedback configuration.
It is shown schematically in figure \ref{wilson2}.
This circuitry guarantees a signal independent feedback resistance.
The feedback resistance can be adjusted by the 
current flowing into transistor N1. 
This current is defined by a current mirror 
(shown dashed in figure \ref{wilson2}).

\psfull

\section{\label{apcmeasure} Measurements of the APC128-DMILL-prototype and the test-structures}

The measurements on the DMILL version of the APC were always accompanied
by the corresponding measurement with the original SACMOS version of the chip 
on the same measuring setup.

All digital parts of the new radiation hard APC128-DMILL worked correctly.
The readout shift register and the pipeline shift register 
have been tested with clock frequencies from $10\, \mbox{kHz}$ 
to $40\, \mbox{MHz}$.

\subsection{Measurements of the preamplifiers \label{measurea}}
All the preamplifier variations  work over a wide range of power consumption
that spans from $\approx 50\, \mu \mbox{W} - 500\, \mu \mbox{W}$.
Unless differently specified our measurements are made with
a power consumption of typically $300\, \mu \mbox{W}$ per channel
which corresponds to $\approx 150\, \mu \mbox{A}$ of supply current.

\begin{figure}[htb]
\begin{center}
\leavevmode
\epsfig{file=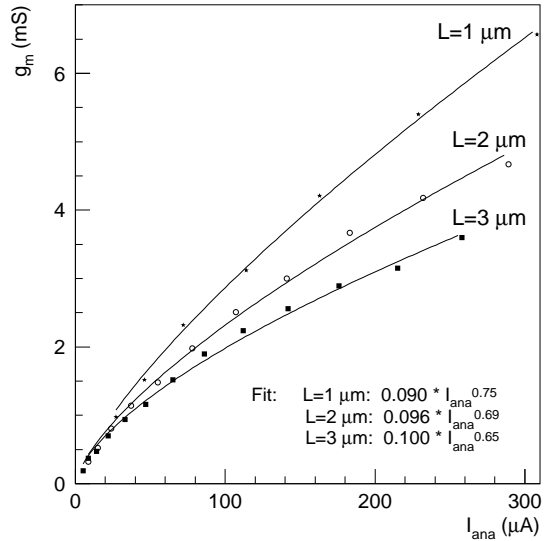,width=7.8cm}
\end{center}
\vspace{-1.cm}
\caption[]{\label{gmplot}Transconductance $g_m$ versus current for different channel length L
measured on a test-structure for the new APC128-DMILL.
A fit of the form $g_m = a  I_{ana}^b$ was performed.}
\end{figure}

\begin{figure}[htb]
\begin{center}
\leavevmode
\epsfig{file=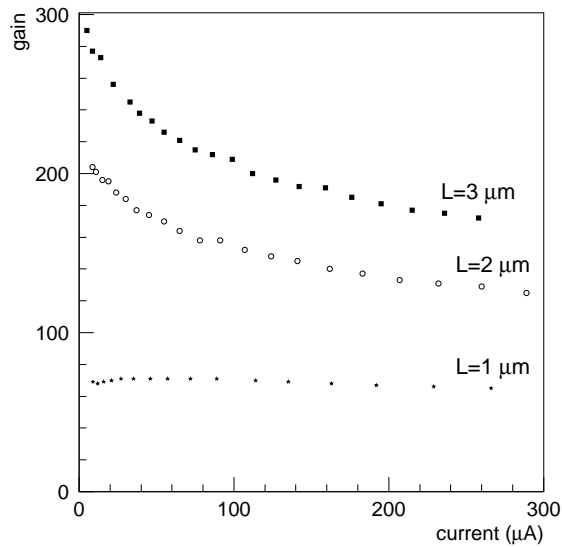,width=7.8cm}
\end{center}
\vspace{-1.cm}
\caption[]{\label{olgplot}Open loop gain
versus current for different channel length L
measured on a test-structure for the new APC128-DMILL}
\end{figure}

The measurement of the  transconductance and the open loop gain as a function 
of the amplifier current is shown in figure \ref{gmplot} 
and \ref{olgplot} respectively. 

The transconductance $g_m$ shows the expected rise with increasing current.
In addition one observes an inverse dependence with the 
channel lengths of the three different amplifier variations.
For our typical operating current of $150\, \mu \mbox{A}$ this gives a 
 transconductance  of 2.5 to $4\, \mbox{mS}$.
The rather large value of  the transconductance of the input transistors
is due to the fact that in this push-pull configuration the $g_m$-values 
of the p-FET and the n-FET are added.
The open loop gain in figure \ref{olgplot} shows in general 
a decreasing behavior as a function of the current.
However, 
for a channel length L=$1\, \mu \mbox{m} $ it appears to be almost flat.
With increasing  channel length we observe a larger open loop gain. 

\psfull
\psfull

As the power density of the white noise is inversely proportional 
to the transconductance ($g_m$) of the input device, 
transconductance preamplifier designs in general tend 
to prefer shorter channel length to obtain a large value of $g_m$.
However, at the same time smaller channel lengths imply   
decreasing open loop gains which limit the obtained Miller 
capacitance of the charge sensitive configuration.
Based on this measurements a compromise channel length of 
$\mbox{L}=2\, \mu \mbox{m}$
appears to be quite reasonable. 

The switch $R12$ (see figure \ref{apcschematic}) allows to use
the preamplifier in the data taking mode with the feedback transistor
connected ($R12$=ON), 
whereas in the pipeline re-read phase the preamplifier is in 
integrating mode ($R12$=OFF).
In the integrating mode we have observed the circuit to work correctly. 
However during data taking mode with the feedback transistor switched on, 
we observe a strong tendency for the amplifier to oscillate.
This was rather surprising since the SACMOS version of the chip 
has proven to be  robust against oscillation.
A detailed analysis of the problem indicates that it is primarily 
a result of the large gate area of the n-FET feedback transistor.
For this feedback transistor a small W/L-ratio
($\mbox{W/L} \approx 1/250$) is desirable since it gives a tolerant 
adjustment of the feedback resistance via $RG$.

The minimal design rules of the DMILL technology  require a larger 
channel width compared to the SACMOS technology which implies a factor 
2.5 in the overall gate area for a identical W/L-ratio.
This results in an increased capacitive load for the conducting channel 
in the feedback transistor. 
The  distributed capacitance to gate and substrate
and the high ohmic resistance 
(few $\mbox{M} \Omega$) of the conducting channel 
form a continuous RC-system. 
This has been identified to be the main source of phase shift 
that allows oscillations to occur.
The feedback path provided by the charge integrating capacitor $C_1$
stabilizes the circuit with increasing value of $C_1$. 
For the DMILL design we measured the parasitic feedback capacitance 
$C_1$ of the preamplifier to be $295 \pm 10\, \mbox{fF}$.
The corresponding measured value of the SACMOS design is $450\, \mbox{fF}$.

As expected the DMILL circuit does not oscillate 
with the additional feedback capacitor $C_2 = 540\, \mbox{fF}$ 
switched on ($CS$=ON).

This allows to operate the prototype chip with the total feedback capacitor 
$C_1 + C_2$  during pipeline write mode
and only $C_1$ during the integrating re-read mode.
This is technically achieved by applying the $IS/SR$ control signal to 
$CS$ as well ($CS=IS/SR$, see figure \ref{apcschematic}).

\subsection{Measuring the two-stage amplifiers \label{twostage}}
\begin{figure}[htb]
\begin{center}
\leavevmode
\epsfig{file=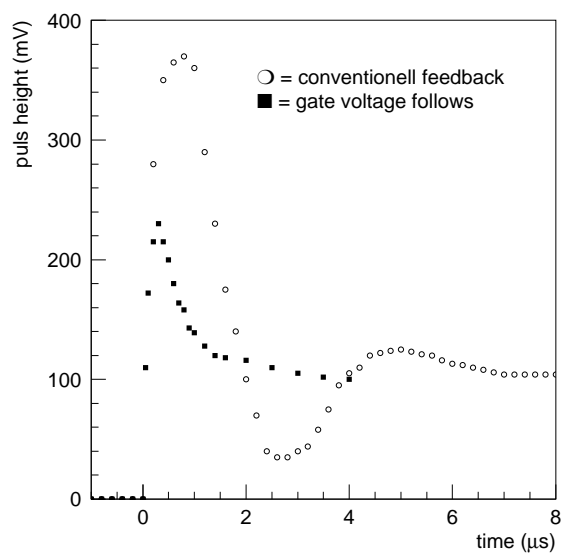,width=7.8cm}
\end{center}
\vspace{-1.cm}
\caption[]{\label{secondstage}Response of the second amplifier stage to
a $100\, \mbox{mV}$ signal step applied to the input.}
\end{figure}

For the detailed understanding of the two stage amplifier chain 
a  separate test-structure containing only the second amplification stage
was fed with a $100 \mbox{mV}$ step function.   
Figure \ref{secondstage} shows the output response of this test structure
for a conventional feedback resistor and the riding feedback configuration 
as described in section \ref{twostagedesign}.
For short pulses the second stage provides an additional gain, 
whereas for DC signals the second stage acts as mere gain 1 follower.
\psfull
The amplifier response with the conventional feedback shows a tendency 
to ring which results in a overshooting gain of 3.7 at its peak position.
In the riding feedback configuration however  this overshoot behavior is 
completely absent. The maximal gain agrees quite well with the designed 
capacitive voltage divider that defines the circuit response for 
fast pulses.
The difference between the observed gain of 2.35 and the 
originally design value of 2 can be understood due to additional parasitic 
capacitances that modify the voltage divider.
For longer time periods both circuits show the expected voltage gain of 1.

\subsection{\label{noise}Noise performance}

For the CST at H1 the most crucial feature of the APC chip is its 
noise performance with a relatively large capacitive load at the input 
of the chip. 
A half-ladder consisting of 3 double-sided sensors 
of the CST represent a capacitive load of $22\, \mbox{pF}$ on the p-side 
and approximately $57\, \mbox{pF}$ on the n-side \cite{cstpaper}. 
This capacitance degrades the quality of the signal.
The charge deposited in the sensor is divided between 
the sensor capacitance and the Miller capacitance of the preamplifier. 
Therefore the bigger the sensor capacitance 
the smaller is the fraction of the charge
effectively collected on the preamplifier.

Therefore we  compared the noise performance of 
different amplifier configurations  
with and without capacitive load at the input.

\psfull
\begin{figure}[htb]
\begin{center}
\leavevmode
\epsfig{file=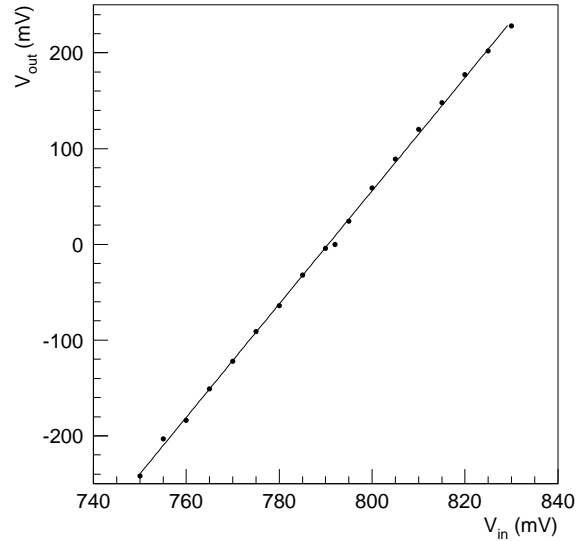,width=7.8cm}
\end{center}
\vspace{-1.cm}
\caption[]{\label{calibration}
Calibration measurement based on the integration of an externally 
applied current to the amplifier input.
Horizontal scale ($V_{in}$): proportional to input current, 
Vertical scale: pulse height at chip output }
\end{figure} 

We measured the variance of the pulse height at the chip output
with no signal present at the input.
This pulse height was converted 
into an equivalent noise charge at the input (ENC).

To calibrate our noise measurements we integrated a small current for 
$2\, \mu \mbox{s}$ on the preamplifier and recorded the corresponding 
pulse height 
at the output of the chip.
The small current was provided by a 
voltage source $V_{in}$ connected via a $5.102\, \mbox{M} \Omega$ resistor 
to the input of the chip.
Hence the calibration constant 
relating a voltage pulse height at the chip output $V_{out}$ to an 
equivalent noise charge at the input 
is $ A_{cal}=\frac {2\, \mu s} {5.1\, M \Omega}
\frac{\Delta V_{in}}{\Delta V_{out}}$, 
with ENC =$ A_{cal} \times V_{out}$. 
A measurement of $V_{out}$ versus $V_{in}$
is shown in figure \ref{calibration}, the shown 
measurement leads to   $ A_{cal}= 414\, e^{-} / \mbox{mV}$.
From this plot the working point of the amplifier can be derived 
by requiring a zero signal height at the chip output.   
For a analog supply voltage of $1.62\, \mbox{V}$ 
an amplifier working point of $790\, \mbox{mV}$ is obtained.

Comparing this method with the  charge signal generated by the on 
chip calibration capacitor a value of $133\, \mbox{fF}$ for the 
calibration capacitor is found
showing good agreement with the design value. 

Several precautions were taken to ensure that the measured noise 
represents the fundamental noise property of the readout chip.
One of them was to operate the measuring setup 
with a complete electric isolation by opto-couplers 
powered  by a regulated  battery supplies in a Faraday cage.
In addition  a differential measurement of two adjacent channels
was performed thus eliminating contributions from pickup that is common 
to all channels.

\begin{figure}[htb]
\begin{center}
\leavevmode
\epsfig{file=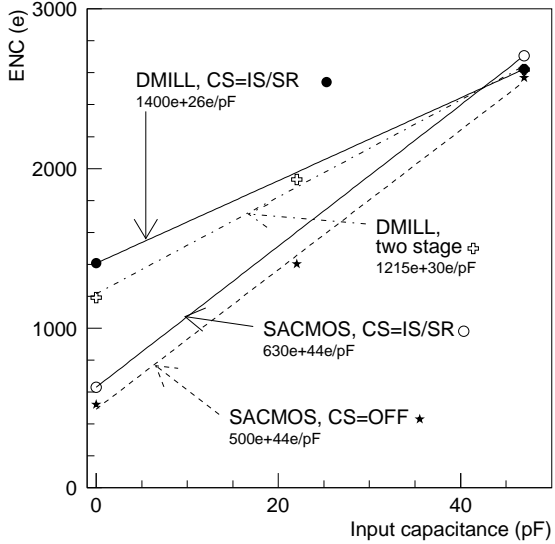,width=7.8cm}
\end{center}
\vspace{-1.cm}
\caption[]{\label{noisestufig}
Measured noise performance (ENC) as a function of the input capacitance
for different operation modes
of the APC chip in SACMOS and DMILL technology
(see text for details).}
\end{figure}

The various  noise measurements had been performed under conditions 
as similarly as possible to the operating conditions of the CST at H1:
The pipeline is running at 10 MHz. 
The noise measurements in figure \ref{noisestufig}
and figure \ref{channellength}
are based on the subtraction of the sum of three pipeline buffers 
before and three buffers after the signal charge is given.

\psfull

Figure \ref{noisestufig} shows the measured equivalent noise charge 
(ENC) of the SACMOS APC and the new DMILL prototype.
The SACMOS chip used in standard H1 operation mode ($CS$=OFF)
shows a noise performance of $500\, e^- + 44 e^-/ \mbox{pF}$  
which is consistent
with previous measurements  \cite{cstpaper}.
  
A direct comparison between the SACMOS and the DMILL chip in this
operation mode was not possible due to the earlier mentioned instability 
of the DMILL preamplifier
with low feedback capacitance.
Therefore we have used an operation mode with $CS$=ON in the sampling phase
and $CS$=OFF in the re-reading phase (i.e. $CS=IS/SR$).
The most remarkable difference between the two chips 
in the observed noise performance is 
in the offset noise value at zero input capacitance.
The slope however shows a better value in the DMILL realization 
compared to the SACMOS such that at large capacitances ($47\, \mbox{pF}$)
the noise  performances are almost equal.

A more detailed study of the zero input capacitance noise
shows that the contribution from the switched on feedback resistance 
($R12$=ON) appears to be much more pronounced in the DMILL chip compared 
to the SACMOS version.
The source of this phenomenon is not fully understood and is 
subject to further study.

The noise performance of a two stage amplifier version 
can also be seen in figure  \ref{noisestufig}.
Since the two stage configuration writes a larger signal
to the pipeline buffers 
noise contribution from the pipeline and the readout system
are reduced.
However the measurements show no significant improvement 
compared to the single stage DMILL version
which  supports our understanding of the extra noise source for 
zero input capacitances being located in the early part 
of the signal amplification stage.

\subsubsection{\label{channelL}Optimizing the channel length}
The noise and speed performance depends on the channel 
length (L) and width (W) of the input transistors. 
Therefore an optimization of this parameters have been done 
by varying these parameters. 

The total channel width of the push pull stage has been kept fixed 
at   $910\, \mu \mbox{m}$ , 
which implies a fixed parasitic feedback capacitances 
of the amplifier variations.

The channel length has been varied a described in section 
\ref{singlestage}.
With a shorter channel length one obtains a higher
transconductance (see figure \ref{gmplot})
which implies a smaller white noise contribution.
The area dependent $1/f$-noise, however, 
is expected to increase. 

The optimal length depend on the optimal ratio of the 
two noise sources and has been determined by 
measuring the noise performance of the amplifiers
with channel length $\mbox{L} = 1,2,3\, \mu \mbox{m}$.

To solely compare the noise directly produced by the input 
transistors we measured the noise behavior of different 
channel length without contribution from the feedback 
transistor ($R12$=OFF) i.e. in integrating mode.
In order to be sensitive to the noise slope
we measured with an capacitor of $47\, \mbox{pF}$ wire 
bonded to the amplifier input.

\psfull
\begin{figure}[htb]
\begin{center}
\leavevmode
\epsfig{file=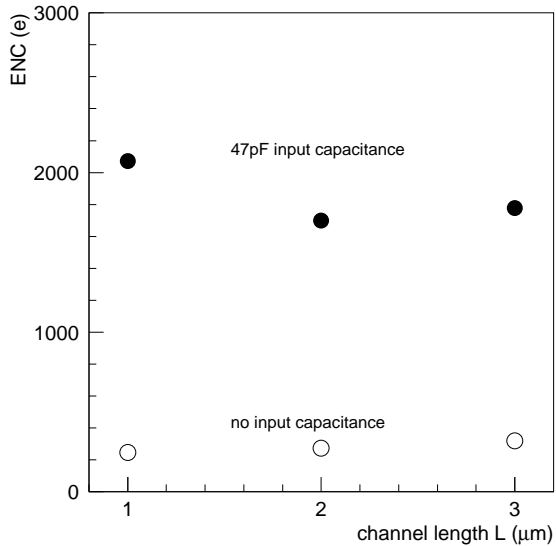,width=7.8cm}
\end{center}
\vspace{-1.cm}
\caption[]{\label{channellength}
Measured ENC for single-stage-amplifier with channel length of 
L=$1\, \mu \mbox{m}$, L=$2\, \mu \mbox{m}$ and L=$3\, \mu \mbox{m}$.}
\end{figure} 
The results are shown in figure \ref{channellength}. 
The measurement clearly shows that the amplifier L=$1\, \mu \mbox{m}$ is 
inferior to the L=$2\, \mu \mbox{m}$ and L=$3\, \mu \mbox{m}$ configurations.
Whereas the L=$2\, \mu$ amplifier is slightly favored
compared to the L=$3\, \mu$ amplifier.
We varied the amplifier current and found that the noise 
dependence on the amplifier current is weak. 

\section{\label{dmilldesign}Final design choices for the APC128-DMILL}

From the measurements presented in section \ref{channelL}
we conclude that the best choice is a channel length
of L=$2\, \mu \mbox{m}$.  

As an increase in channel width is expected to improve the noise 
behavior and to increase the basic feedback capacitance
we increased the channel length compared to the prototype 
by approximately 20 per cent.

Apart from the feedback $C_2$ with a value of $300\, \mbox{fF}$
that can be activated by the switch $CS$=ON
a new additional feedback capacitor $C_2'$ 
with a value of $150\, \mbox{fF}$ was added.
This allows the user to choose
between four different 
feedback capacitance values.

For the feedback resistance we replaced the long n-MOS-feedback-transistor 
by a riding feedback circuit as described in section \ref{twostagedesign}.
Our measurements of the two stage amplifier (see section \ref{twostage})
with the riding feedback circuit shows strong evidence 
for an improved robustness against oscillation
which is also supported by simulations \cite{simulation}.

Since the two stage amplifier configuration did not show 
any convincing improvement in noise performance
we realized the single stage architecture, 
as used in the old SACMOS chip
The single stage solution gives maximum compatibility
to the SACMOS APC and smallest power dissipation .

All other building blocks of the prototype were found to work well 
and were therefore used  unaltered in the final design.

\section*{Acknowledgment}
We wish to thank Silvan Streuli for his substantial help.
For encouraging discussion we are very grateful to
Ralph Eichler and Daniel Pitzl.

\end{document}